# First observation of magnetic flux rope inside electron diffusion region

Z. Z. Chen[1,2], H. S. Fu[1,2,†], Z. Wang[1,2], Z. Z. Guo[1,2], Y. Xu[1,2], C. M. Liu[1,2]

[1]School of Space and Environment, Beihang University, Beijing, China

[2]Key Laboratory of Space Environment Monitoring and Information Processing, Ministry of Industry and Information Technology, Beijing, China

†Correspondence to H. S. Fu at huishanf@gmail.com

**Key points:**

- We present the first observation of a magnetic flux rope (MFR) inside an electron diffusion region (EDR).

- The MFR, with a width of ~27.5 $d_e$ in the L direction and ~4.8 $d_e$ in the N direction, was moving away from the X-line.

- We reconstructed magnetic topology of the electron-scale MFR.

**Abstract.** Magnetic flux ropes (MFRs) play a crucial role during magnetic reconnection. They are believed to be primarily generated by tearing mode instabilities in the electron diffusion region (EDR). However, they have never been observed inside the EDR. Here, we present the first observation of an MFR inside an EDR. The bifurcated non-force-free MFR, with a width of ~27.5 $d_e$ in the L direction and ~4.8 $d_e$ in the N direction, was moving away from the X-line. Inside the MFR, strong energy dissipation was detected. The MFR can modulate the electric field in the EDR. We reconstructed magnetic topology of the electron-scale MFR. Our study promotes understanding of MRFs' initial state and its role in electron-scale processes during magnetic reconnection.

# 1. Introduction

Magnetic reconnection, a fundamental plasma process, is widely believed to be responsible for many explosive energy release phenomena in astrophysical plasmas, e.g. coronal mass ejection (*Chen & Shibata, 2000*) and terrestrial substorms (*Angelopoulos et al., 2008; Cao et al., 2010, 2013; Fu et al., 2012, 2013a; Xu et al., 2018; Wei et al., 2007*), and in the disruptions of fusion experiments (*Ji et al., 2008*). It is believed to begin in the electron diffusion region (EDR), where electron demagnetization and energy dissipation occur (*Øieroset et al., 2001; Burch et al., 2016a; Cao et al., 2017; Norgren et al., 2016*). The EDR is embedded in the ion diffusion region (IDR; *Øieroset et al., 2001; Fu et al., 2019a*) because of different motions between ions and electrons. Magnetic reconnection can lead to reconfiguration of magnetic field topology (*Chen et al., 2019a; Fu et al., 2019b, 2020a*) and formation of many structures, such as reconnection fronts (*Fu et al., 2013b, 2019c, 2020b; Liu et al., 2018*), magnetic nulls (*Chen et al., 2018; Fu et al., 2020a*), and magnetic flux ropes (MFRs; also called magnetic islands or plasmoids; *Slavin et al., 2003; Deng et al., 2004; Daughton et al., 2011; Drake et al., 2006a; Eastwood et al., 2016; Huang et al., 2016a; Chen et al., 2019b, 2019c; Fu et al., 2016*), which, in turn, play crucial roles in magnetic reconnection (*Daughton et al., 2011; Huang et al., 2012; Oka et al., 2010; Wang et al., 2015*).

Simulations have demonstrated that the majority of MFRs are produced by tearing mode instabilities inside the EDR (*Daughton et al., 2011; Drake et al., 2006a; Lu et al., 2019*). Spontaneously generated, such MFRs evolve and interact with each other, leading to turbulence (*Daughton et al., 2011; Chen et al., 2020; Cozzani et al., 2019; Fu et al., 2017; Wang et al., 2015*), where electron-scale physical processes play crucial roles (*Daughton et al., 2011*). Meanwhile, the MFRs are ejected out from the EDR after growing for some period (*Daughton et al., 2006*). They can break the reconnection current sheet, causing the dramatic enhancement of the reconnection rate (*Daughton et al., 2006, 2009*). In addition, during magnetic reconnection, they contribute greatly to electron acceleration, involving coalescence of MFRs (*Retinò et al., 2007; Wang et al., 2015; Zhou et al., 2017*), contraction of MFRs (*Chen et al., 2007; Drake et al., 2006b; Huang et al., 2012; Retinò et al., 2008*), and reconnection electric field when electrons are trapped in MFRs (*Oka et al., 2010*).

MFRs have been widely observed in the solar heliosphere (*Song et al., 2012; Zhang et al., 2012; Kumar et al., 2019*), terrestrial magnetosphere (*Wang et al., 2010; Fu et al., 2017; Khotyaintsev et al., 2010; Huang et al., 2012, 2016a; Eastwood et al., 2016; Hwang et al., 2018; Chen et al., 2019b*), and laboratory experiments (*Dorfman et al., 2012*). These observed MFRs were far from the X-line, either in IDR or exhaust region outside the diffusion region. Recently, a series of MFRs were detected near the EDR (*Huang et al., 2018*). However, MRFs have never been observed inside the EDR so far, owing to a lack of high-resolution observation data and the electron-scale of MFRs inside the EDR. The terrestrial magnetotail, with a larger electron inertial length (the order of 20 km), is a natural laboratory to investigate the existence of MFRs inside the EDR. In this study, we observe for the first time an MFR inside an EDR and investigate its properties, utilizing high-resolution data obtained from the

Magnetospheric Multiscale (MMS) mission (*Burch et al., 2016b*).

## 2. Observations

The observation data are from the MMS mission (*Burch et al., 2016b*). Specifically, the magnetic field data (8-ms resolution) are recorded by the Fluxgate Magnetometer (FGM; *Russell et al., 2016*); the electric field data (8192 Hz resolution) are collected by the Axial Double Probe (ADP; *Ergun et al., 2016*) and the Electric Double Probe (EDP; *Lindqvist et al., 2016*); the plasma data (150-ms resolution for the ions and 30-ms resolution for the electrons) are measured by the Fast Plasma Investigation Instrument (FPI; *Pollock et al., 2016*). Throughout the paper, Geocentric Solar Magnetospheric (GSM) coordinates are used unless otherwise specified.

The event of interest was observed on 27 August 2018, 12:15:30 UT to 12:16:30 UT, when the spacecraft were located at [-21.07 13.12 0.97] $R_E$ (the Earth's radius) in GSM coordinates with separations of ~35 km (~1.5 $d_e$; $d_e = c/\omega_{pe} \approx 5.31 N_e^{-0.5}$ is the local electron inertial length, $N_e$~0.05 cm$^{-3}$). Figures 1b-1f present an overview of the event. These data were recorded by MMS1 in the burst mode (*Burch et al., 2016b*). During the interval, MMS measured small magnetic field $B_x$ (|$B_x$|< 10 nT; Figure 1b) and large plasma β (β > 0.5; Figure 1f), indicating that it was in the terrestrial plasma sheet (*Cao et al., 2006*). As can be seen, the magnetic field $B_x$ component changed from negative (southern hemisphere) to positive (northern hemisphere) at ~12:15:43 UT, then changed from positive (northern hemisphere) to negative (southern hemisphere) at ~12:16:10 UT, meaning that MMS crossed the plasma current sheet twice. At ~12:15:43 UT and ~12:16:10 UT, the magnetic field strength reached ~0 nT and ~2 nT, respectively (Figure 1b), the magnetic field $B_x$ component reversed (Figure 1b), and the plasma β sharply increased (Figure 1f). These features suggest the crossing of the central current sheet. During this period, MMS detected tailward ion flow ($V_{ix}$<0) before ~12:15:55 UT and earthward ion flow ($V_{ix}$>0) after ~12:15:55 UT (see blue curve Figure 1d). Coincident with the ion flow reversal, the magnetic field $B_z$ component also reversed roughly from negative to positive (Figure 1b), suggesting ongoing magnetic reconnection (*Burch et al., 2016a; Ergun et al., 2018; Huang et al., 2018; Øieroset et al., 2001; Torbert et al., 2018*). Figure 1a shows the MMS trajectory (blue arrowed line). MMS firstly crossed the tailward outflow, then moved in the inflow region, and finally crossed the earthward outflow. Therefore, the reversal of ion flow $V_x$ component was detected between two reversals of the magnetic field $B_x$ component. The first crossing of the current sheet (~12:15:43 UT) lasted about 8 seconds, while the second one (~ 12:16:10 UT) lasted 10 seconds. Based on the timing analysis on magnetic field $B_x$ component, the width of the first current sheet crossing is estimated as ~225 km (~0.31$d_i$; $d_i = c/\omega_{pi} \approx 228 N_i^{-0.5}$ is the local ion inertial length, $N_i$~0.1 cm$^{-3}$), while the width of the second current sheet crossing is estimated ~460 km (~0.78$d_i$, $N_i$~0.15 cm$^{-3}$). It indicates that the first crossing is closer to the X-line.

During the first crossing of the reconnection current sheet, MMS detected a bipolar variation of the electric field $E_z$ component from 30 mV/m to -20 mV/m (see red curve in Figure 1c), which can be interpreted as the Hall electric field in the two-fluid model of reconnection (*Øieroset et al., 2001; Peng et al., 2017*). In addition, MMS detected tailward ($V_{ex}$<0) and dawnward ($V_{ey}$<0) electron jet (Figure 1e). Therefore, MMS may have crossed an EDR. A small bipolar variation of magnetic field $B_z$ component was detected at ~12:15:44.75 UT in the electron jet (see red curve in Figure 1b). The bipolar variation of the $B_z$ component matches the features of magnetic flux ropes (MFRs; *Huang et al., 2016a; Wang et al., 2010; Chen et al., 2019c*). These features indicate the possibility that an MFR was observed in an EDR. During the second crossing of the reconnection current sheet, MMS did not detect coincident Hall magnetic field and electric field, meaning that MMS crossed downstream of the diffusion region.

In Figure 2, we investigate whether an MFR was observed in the EDR. Firstly, encounter of an EDR should be demonstrated. Here we focus on the first crossing of the reconnection current sheet, i.e., 12:15:30 to 12:15:55. During this period, magnetic field data collected by all four spacecraft in burst mode, and plasma data and electric field data measured by MMS1 in burst mode are used. To describe the reconnection current sheet, we performed the hybrid minimum variance analysis (MVA) (*Phan et al., 2018*) on the magnetic field during 12:15:40–12:15:47 UT to obtain the LMN coordinate. With respect to the original GSM coordinates, L=[0.90 -0.43 0.07], M=[0.43 0.87 -0.22], and N=[0.03 0.23 0.97]. The LMN coordinate system is close to the GSM coordinate system.

During the crossing of the reconnection current sheet (reflected by the variation of $B_L$ component from -6 to 6 nT that was detected by all spacecraft; Figure 2a), MMS1 detected an intense electron jet (Figure 2d) with $V_{eL}$ ~ 27 $V_{A,in}$ and $V_{eM}$ ~ 54 $V_{A,in}$ ($V_{A,in} = B/\sqrt{\mu_0 m_i N_i}$ ~ 550 km/s is the Alfvén speed, B~8nT, $N_i$~0.1cm$^{-3}$). The timing analysis on magnetic field $B_L$ component during the electron jet (12:15:39.00-

12:15:46.70 UT) indicates that the width of the electron current layer is ~210 km (~9 $d_e$). The width of the electron current layer obtained by using Ampere's law (*Torbert et al., 2018*) is 80 km (~3.3 de), much different from the results of timing analysis. It may be because that the assumption ($B_N \ll B_L$) is not applicable in our event, where $B_N$ and $B_L$ are on the same order of magnitude. The width of the electron current layer may be ~3.3$d_e$ – ~9$d_e$. The electron current layer has a bifurcated structure (*Liu et al., 2013*), with dip of electron velocity $V_{eL}$ component and $V_{eM}$ component near the reversal of magnetic field $B_L$ component (Figures 2a and 2d). The electron bulk velocity was about 70 times larger than the ion bulk velocity (Figures 1d-1e), which leads to an intense electron-driven current. Inside the electron jet, the measured electric field and electron convection term did not match (**E**+**V**$_e$×**B**≠**0**; Figure 2e), meaning that electrons were demagnetized. The electron-driven current and the electron demagnetization usually lead to strong energy dissipation. Figure 2f shows (**J**•(**E**+**V**$_e$×**B**)), which is obtained in the electron rest frame (*Zenitani et al., 2011*). Here the current density is derived from particle moments (**J**=$N_i$·e·**V**$_i$-$N_e$·e·**V**$_e$). The pink shading represents errors of **J**•(**E**+**V**$_e$×**B**), where both plasma errors and electric field errors are included. As can be seen, **J**•(**E**+**V**$_e$×**B**) with value from -0.5 to 0.5 nW/m$^3$ was detected inside the electron jet. The positive **J**•(**E**+**V**$_e$×**B**) indicates energy dissipation that electromagnetic field energy is converted to plasmas energy, while the negative **J**•(**E**+**V**$_e$×**B**) suggests that plasmas energy is converted to the electromagnetic field energy.

To quantify the electron nongyrotropy (*Swisdak 2016*), we calculated the $\sqrt{Q}$ by using the electron pressure tensor. Inside the electron jet, sharp enhancement of electron nongyrotropy was observed (Figure 2g). The value of $\sqrt{Q}$ reached ~0.6, much larger than the background value. Such enhancements of $\sqrt{Q}$ indicate the existence of crescent distributions (*Norgren, et al., 2016*). Figures 2h-2m present the electron distribution functions in the plane perpendicular to the local magnetic field during the electron jet. The magenta vertical bars at the top of Figure 2a mark when the electron distribution functions were collected. In Figures 2h-2m, clear crescent distribution can be seen in the half plane of $V_{(e \times b)} > 0$, where e and b represent unit vector of electric field and magnetic field, respectively.

When the magnetic field $B_L$ component reversed, the magnetic field $B_M$ component and $B_N$ component were ~0.5 nT and ~0 nT, respectively. It suggests the existence of magnetic nulls. We apply the First-Order Taylor Expansion (FOTE) method (*Fu et al., 2015, 2016, 2020a*) near the reversal of magnetic field $B_L$ component (12:15:42.97-12:15:43.08 UT, yellow shading in Figures 2a-2g) to examine whether magnetic nulls exist inside the electron jet. Figures 2n-2p presents the FOTE results, including (panel n) the null-MMS distance, (panel o) the minimum null-MMS distance, and (panel p) the two parameters, $\eta \equiv |\nabla \cdot \mathbf{B}|/|\nabla \times \mathbf{B}|$ and $\xi \equiv |\lambda_1 + \lambda_2 + \lambda_3|/|\lambda|_{max}$, for quantifying the quality of the FOTE results. A 3-D magnetic null can be classified into two types, radial type and spiral type. The radial-type null includes A null and B null, while the spiral-type null includes As null and Bs null. In some conditions, the 3-D radial null and spiral null can degenerate into 2-D X-line and O-line, respectively (*see Fu et al., 2015, 2020a for details*). During the interval (yellow shading), the X-line was found (Figure 2o). As can be seen, the null-MMS distance is quite small (Figures 2n-

2o), with the minimum distance of ~10 km (~0.4 $d_e$; Figure 2o). Since one of the eigenvalues of the Jacobian matrix $\nabla B$ is very small, the radial nulls degenerate to an X-line (Figure 2o; *see Fu et al., 2015 for details*). The two parameters, η and ξ, are quite small (<40%; Figure 2p), indicating that the structure around MMS during this short interval is generally linear, and thus the FOTE results are reliable.

We apply the FOTE method to reconstruct the magnetic topology of the X-line, as done in previous studies (*Fu et al., 2016, 2017, 2019b; Chen et al., 2019a*). For better revealing the topological features of X-line, the eigenvector coordinates ($e_1$, $e_2$, $e_3$) obtained from the Jacobian matrix $\nabla B$ are used (*Fu et al., 2015, 2019b; Liu et al., 2019a; Wang et al., 2019a; Chen et al., 2019a*). Figure 2q shows the 2D topology of the X-line. We find an angle of ~37° between the two separatrix surfaces, indicating fast reconnection with reconnection rate of 0.20 (*Liu et al., 2017*). Here, the reconnection rate is calculated by using $R = \tan(\frac{\theta}{2})(\frac{1-tan^2(\frac{\theta}{2})}{1+tan^2(\frac{\theta}{2})})^2 \sqrt{1-tan^2(\frac{\theta}{2})}$, where θ is the angle between the two separatrix surfaces. The reconnection rate calculated by using $\frac{E_r}{V_{Ai}B}$ is ~0.18 ($E_r$~0.8 mV/m is the average value of $E_M$ in the electron jet, $V_{Ai}$~550 km/s, B~8 nT), which is consistent with the result by using opening angle.

These features, including electron demagnetization, energy dissipation, electron nongyrotropy (crescent distribution), and an X-line, indicate that MMS encountered an EDR (*Burch et al., 2016a; Fu et al., 2019b; Chen et al., 2019a; Wang et al., 2019b*) in the region marked by grey shading in Figures 2a-2g. In the EDR, a small bipolar variation of the magnetic field $B_N$ component from ~1 nT to ~-2.8 nT was detected at

~12:15:44.75 UT (marked by cyan shading; Figure 2c), which is widely believed to be a signature of MFRs (*Huang et al., 2016a; Wang et al., 2010; Chen et al., 2019c; Liu et al., 2019b*). Therefore, an MFR may be observed inside an EDR.

To confirm the encounter of an MFR, the spiral feature of magnetic field lines should be captured. We use the FOTE method to reconstruct magnetic topology of the structure, as done in previous studies (*Chen et al., 2019b, 2019c; Fu et al., 2016; Wang et al., 2017*). For better revealing the topological features of the magnetic field, the LMN coordinates is used. Figure 3a and 3b shows the 3D topology of the structure at different view angles. The color scale in Figure 3 denotes the magnetic field strength |B| and the blue arrows on the lines denote the field direction. Grey lines denote grids. As can be seen, the magnetic field topology shows spiral feature, indicating an encounter of an MFR. The axis of the MFR is roughly along the M direction (Figure 3a). The topology in Figure 3b is roughly the projection of the MFR in the LN plane. As can be seen, positive-to-negative bipolar variation of $B_N$ are observed when MMS moves along L direction. Therefore, an MFR is indeed observed inside an EDR.

Now we focus on the properties of the MFR. The positive-to-negative bipolar variation of $B_N$ (Figure 4c) indicates that MMS crossed the MFR roughly along the L (close to $X_{GSM}$) direction (see the MMS trajectory marked by blue arrowed line in Figure 1a). Inside the MFR, the magnitude of $B_M$ component decreased (Figure 4b), suggesting the MFR is categorized as a bifurcated MFR as discussed in previous studies (*Zhang et al., 2010*). We perform timing analysis on magnetic field $B_N$ component for the crossing of the MFR (marked by cyan shading in Figures 4a-4i, i.e. 12:15:44.528-

12:15:44.985 UT) and obtain that the speed of the MFR was 2,200 × [-0.69, -0.72, -0.12] km/s in the LMN coordinate system, suggesting that the MFR was moving away from the X-line. Hence, the width of the MFR in the L direction and N direction are ~690 km (~27.5 $d_e$) and ~120 km (~4.8 $d_e$), respectively. The electron-scale MFR was elongated, maybe because that it was generated in an elongated EDR. The speed of the MFR matches the trajectory reflected by the positive-to-negative bipolar variation of $B_N$ (Figure 4c).

As can be seen, positive **J**•(**E**+**V**$_e$×**B**) with value up to ~0.3 nW/m$^3$ inside the MFR (Figure 2f) suggests energy dissipation that electromagnetic field energy is converted to plasmas energy, which was also reported inside an ion-scale MFR (*Huang et al., 2019*). Energy dissipation inside the MFR is stronger than that in adjacent regions (Figure 2f), suggesting energy dissipation inside the MFR may be related to the MFR physics. The energy dissipation indicates that there are still electron-scale physical processes inside the MFR, namely that the MFR was still dynamical. In addition, clear crescent distribution in the half plane of $V_{(e×b)}$ > 0 was detected (Figure 2m). However, the electron crescent inside the MFR was less distinct than that in other regions of the EDR. Adjacent to the trailing edge of the MFR, MMS observed an intense parallel electric field (Figure 4d), which is interpreted as electrostatic solitary waves (ESWs; *Fu et al., 2020c; Khotyaintsev et al., 2020*). Figures 4j-4k show zoomed-in plot of the ESWs measured by MMS1 and MMS4, respectively. They are characterized by bipolar variation of parallel electric filed (Figures 4j-4k). The ESWs were also detected inside and near ion-scale flux ropes (*Huang et al., 2016b, 2019*). Inside the MFR, positive-to-negative bipolar variation of electric field $E_L$ component (black curve in Figure 4e) was detected. As can be seen, the Hall term (blue curve in Figure 4e) make great contributions to balance the electric field $E_L$ component. *Zhou et al. (2012)* found bipolar variation of electric field $E_x$ component in simulation, where the electric field $E_x$ component was primarily balanced by the Hall term. It is different from our observation, where other terms in the generalized Ohm's law (e.g. the divergence of electron pressure tensor) cannot be ignored at the peak (~12:15:44.650) and dip (~12:15:44.900) of $B_N$. However, we cannot calculate other terms due to a lack of electron data measured by MMS4. The Hall term in L direction is dominated by $J_M B_N$, because that the electron-driven current is primarily along the M direction (Figure 2d), namely $J_M \gg J_N$. Therefore, the combination of reconnection current ($J_M$) and bipolar variation of $B_N$ leads to the formation of bipolar variation of electric field $E_L$ component. In other word, the MFR can modulate the electric field. In the M direction and the N direction, the Hall term also make great contributions to balance the electric field (Figures 4f-4g). Throughout the MFR, the perpendicular current was larger than the parallel current (Figure 4h), indicating it was a non-force-free structure. Inside the MFR, both the perpendicular current and the plasma thermal pressure increased (Figures 4h-4i).

## 3. Conclusions and Discussions

We have presented the first observation of a magnetic flux rope (MFR) in an electron diffusion region (EDR). Electron demagnetization, energy dissipation, electron nongyrotropy (crescent distribution) and an X-line were all observed, indicating an

EDR encounter. The MFR, with a width of ~27.5 $d_e$ in the L direction and ~4.8 $d_e$ in the N direction, was moving away from the X-line. The $B_M$ component decreased at the center of the MFR, suggesting that the MFR is categorized as a bifurcated MFR. Throughout the MFR, the perpendicular current was larger than the parallel current, indicating that the MFR is a non-force-free structure. Inside the MFR, MMS detected strong energy dissipation from the electromagnetic field to the particles, indicating that the MFR was still dynamical. The bipolar variation of the electric field $E_L$ component suggests that the MFR can modulate the electric field in the EDR. In addition, adjacent to the trailing edge of the MFR, electrostatic solitary waves (ESWs) were observed by MMS. We reconstructed the magnetic topology of the electron-scale MFR. The electron-scale MFR detected inside the EDR suggest that the MFR is in initial stage of MFR growth. Our study provides insights into MRFs' initial state and its role in electron-scale processes during magnetic reconnection.

The MFRs can make great contributions to electron energization during magnetic field (*Chen et al., 2007; Drake et al., 2006b; Huang et al., 2012; Oka et al., 2010; Zhou et al., 2017*). However, compared with the other regions in the EDR, electron acceleration/heating was not obviously observed inside the MFR. The electron energization mechanisms proposed by previous studies include (1) adiabatic acceleration (*Drake et al., 2006b; Huang et al., 2012*), (2) reconnection electric field (*Oka et al., 2010*), and (3) coalescence of MFRs (*Zhou et al., 2017*). Since electrons are demagnetized inside the EDR, the adiabatic condition is broken. Electrons trapped by electrostatic potential inside MFRs can be accelerated by reconnection electric field

(*Oka et al., 2010*). However, the convergent electric field $E_L$ associated the MFR (Figure 4e) lead to a negative potential that cannot trap electrons inside the MFR. In addition, there is no evidence of coalescence of MFRs. Therefore, these three mechanisms cannot work in the electron-scale MFR inside the EDR. More studies are needed to investigate whether such electron-scale MFR can play roles in electron energization inside the EDR and which mechanism works.

The observation of MFR inside the EDR suggests possible existence of electron-scale turbulence inside the EDR, namely that the EDR has inhomogeneous structure at electron scale (*Cozzani et al., 2019*). Inside the EDR, MMS observed intermittent energy dissipation (Figure 2f), which seems to be the consequence of the filament current driven by electrons (Figure 2d). Inside the MFR, energy dissipation up to ~0.3 nW/m$^3$ was detected (Figure 2f). The observation of MFR inside the EDR also suggests the possible existence of multiple X-lines inside the EDR. During the formation of new X-line, the energy dissipation and conversion occur. Therefore, the formation of MFR may elevate rate of energy dissipation and conversion during magnetic reconnection.

## Acknowledgments

The data used is available from the MMS Science Data center (https://lasp.colorado.edu/mms/sdc/public/). This work was supported by NSFC grants 41821003 and 41874188 and the ISSI travel grant for team "MMS and Cluster Observations of Magnetic Reconnection".

# Figure captions

**Figure 1.** Overview of the magnetic reconnection encounter. (a) Schematic of the magnetic reconnection and satellite trajectory. MMS1 observations of (b) magnetic field, (c) electric field, (d) ion bulk velocity, (e) electron bulk velocity, and (f) plasma β. These data were collected in burst mode and shown in Geocentric Solar Magnetospheric (GSM) coordinates. Magnetospheric Multiscale mission=MMS.

**Figure 2.** A magnetic flux rope (MFR) observed in an electron diffusion region (EDR). MMS observations of magnetic field (a) L component, (b) M component, and (c) N component in LMN coordinates. With respect to the original GSM coordinates, L = [0.90 -0.43 0.07], M = [0.43 0.87 -0.22], and N = [0.03 0.23 0.97]. MMS1 observations of (d) electron velocity in LMN coordinates, (e) electric field in the electron rest frame ($\mathbf{E}' = \mathbf{E} + \mathbf{V_e} \times \mathbf{B}$), (f) energy dissipation ($\mathbf{J} \cdot \mathbf{E}'$, black curve) with errors (pink shading) from both plasma and electric field, and (g) electron nongyrotropy ($\sqrt{Q}$). MMS1 observations of electron distribution functions at (h) 12:15:41.076, (i) 12:15:41.646, (j) 12:15:42.006, (k) 12:15:42.936, (l) 12:15:43.656, (m) 12:15:44.766 in the panel perpendicular to local magnetic field ($V_{e \times b} - V_{b \times (e \times b)}$), where b and e represent the unit vector of magnetic field and electric field, respectively. The electron distributions were integrated from $-10^5$ km/s to $10^5$ km/s in the out-of-plane direction. To eliminate the effects of photoelectrons, the data at energy below 100 eV are excluded, as done in *Xu et al. (2019)*. (n-q) First-Order Taylor Expansion (FOTE) results performed during the interval marked by yellow shading (12:15:43.000-12:15:43.140 UT) in Panels a-g. (n) The null-MMS distance, (o) the minimum null-MMS distance, (p) two parameters, $\eta \equiv |\nabla \cdot \mathbf{B}|/|\nabla \times \mathbf{B}|$ (black) and $\xi \equiv |\lambda_1 + \lambda_2 + \lambda_3|/|\lambda|_{max}$ (blue), for quantifying the quality of the FOTE results, and (q) 2D view of reconstructed topology of the X-line. The blue arrows denote the magnetic field directions, and the color scale denotes the magnetic field strength. The MFR is marked by cyan shading (12:15:44.528-12:15:44.985 UT) in Panels a-g. The EDR is marked by grey shading (12:15:39.00-12:15:46.70 UT) in Panels a-g.

**Figure 3.** (a-b) 3D magnetic topology of the magnetic flux rope (MFR) reconstructed by using FOTE method at different view angles. The blue arrows denote the magnetic field directions, and the color scale denotes the magnetic field strength. Grey lines denote grids.

**Figure 4.** Properties of the magnetic flux rope (MFR). MMS observations of magnetic field (a) L component, (b) M component, and (c) N component in LMN coordinates. (d) MMS observations of parallel electric field. (e-g) Electric field measured by MMS1 (black curve) and ion convection term (red curve) and Hall term (blue curve) in the generalized Ohm's law in (e) the L direction, (f) the M direction and (g) the N direction. MMS1 observations of (h) perpendicular (black curve) and parallel (red curve) current, (i) magnetic pressure (red curve), plasma thermal pressure (blue curve), and total pressure (black curve). Here, we increase the resolution of plasma data to the resolution of magnetic field by performing linear interpolation. (j-k) Parallel electric field

measured by (j) MMS1 and (k) MMS4 during 12:15:45.15-12:15:45.31 UT. The MFR is marked by cyan shading (12:15:44.528-12:15:44.985 UT) in Panels a-i.

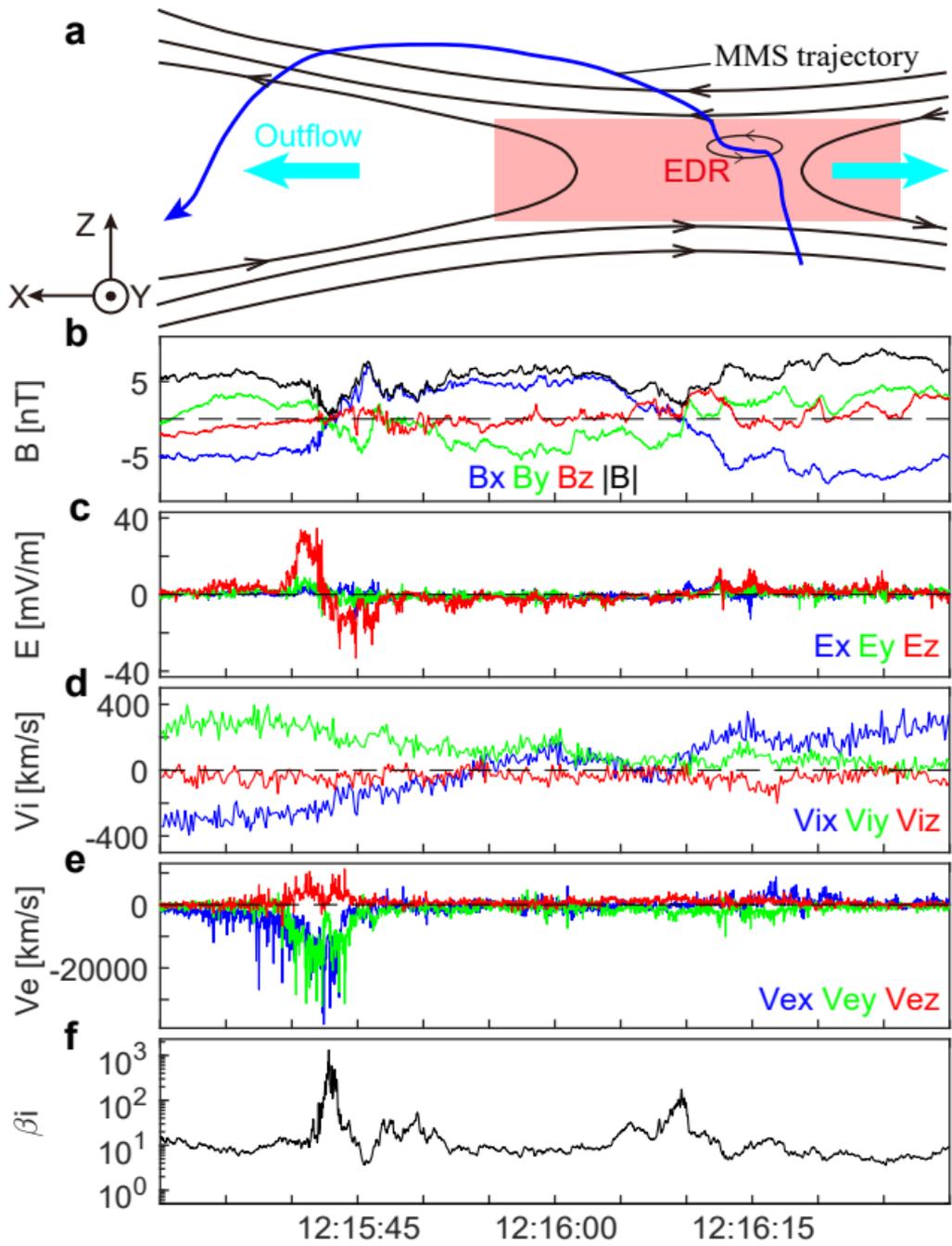

**a** MMS trajectory
Outflow
EDR
Z
X ⊙ Y

**b** B [nT] — Bx By Bz |B|
**c** E [mV/m] — Ex Ey Ez
**d** Vi [km/s] — Vix Viy Viz
**e** Ve [km/s] — Vex Vey Vez
**f** βi

12:15:45   12:16:00   12:16:15

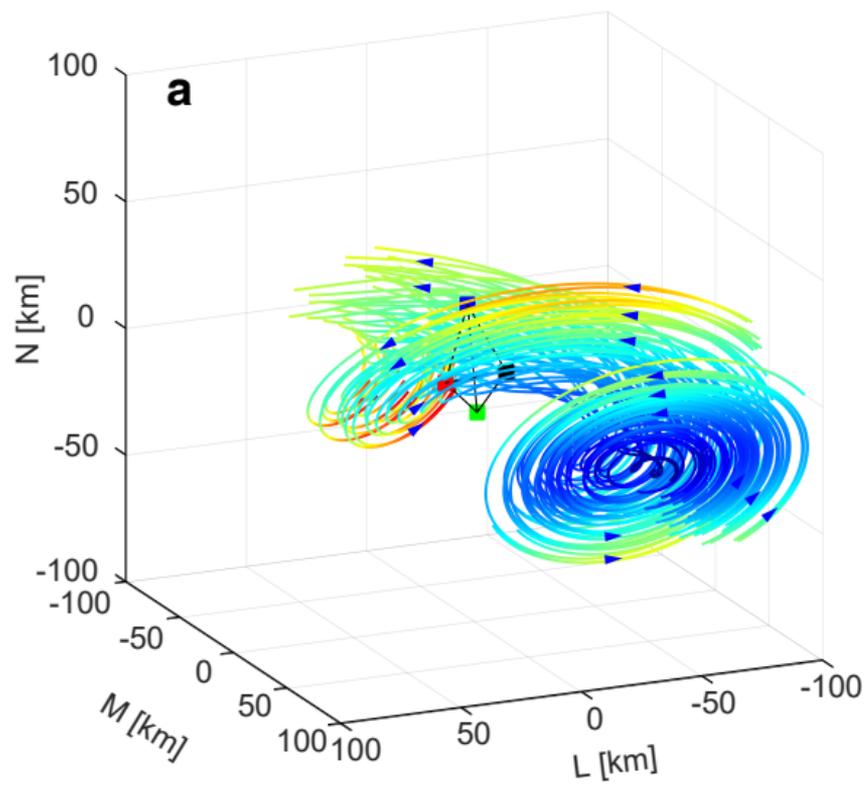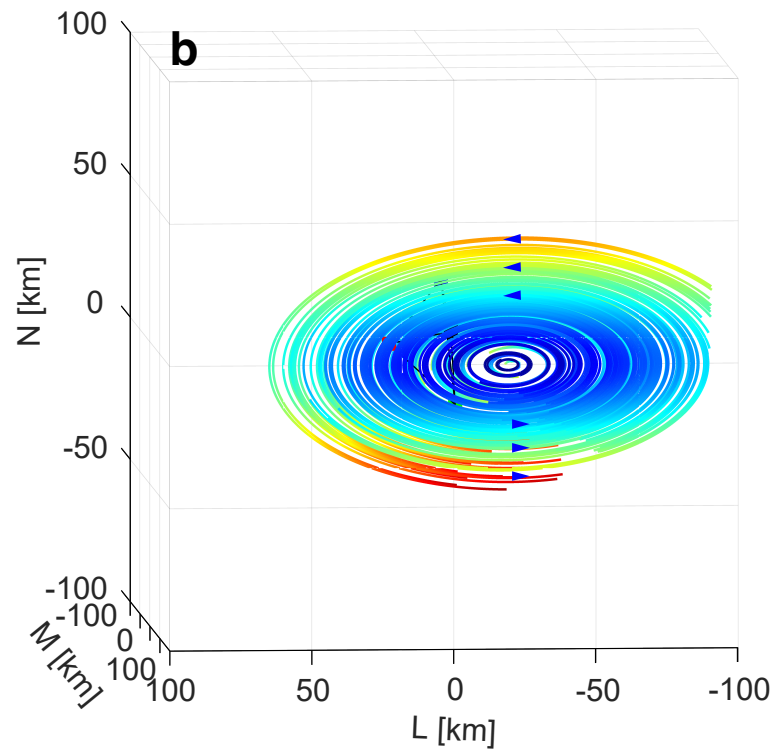

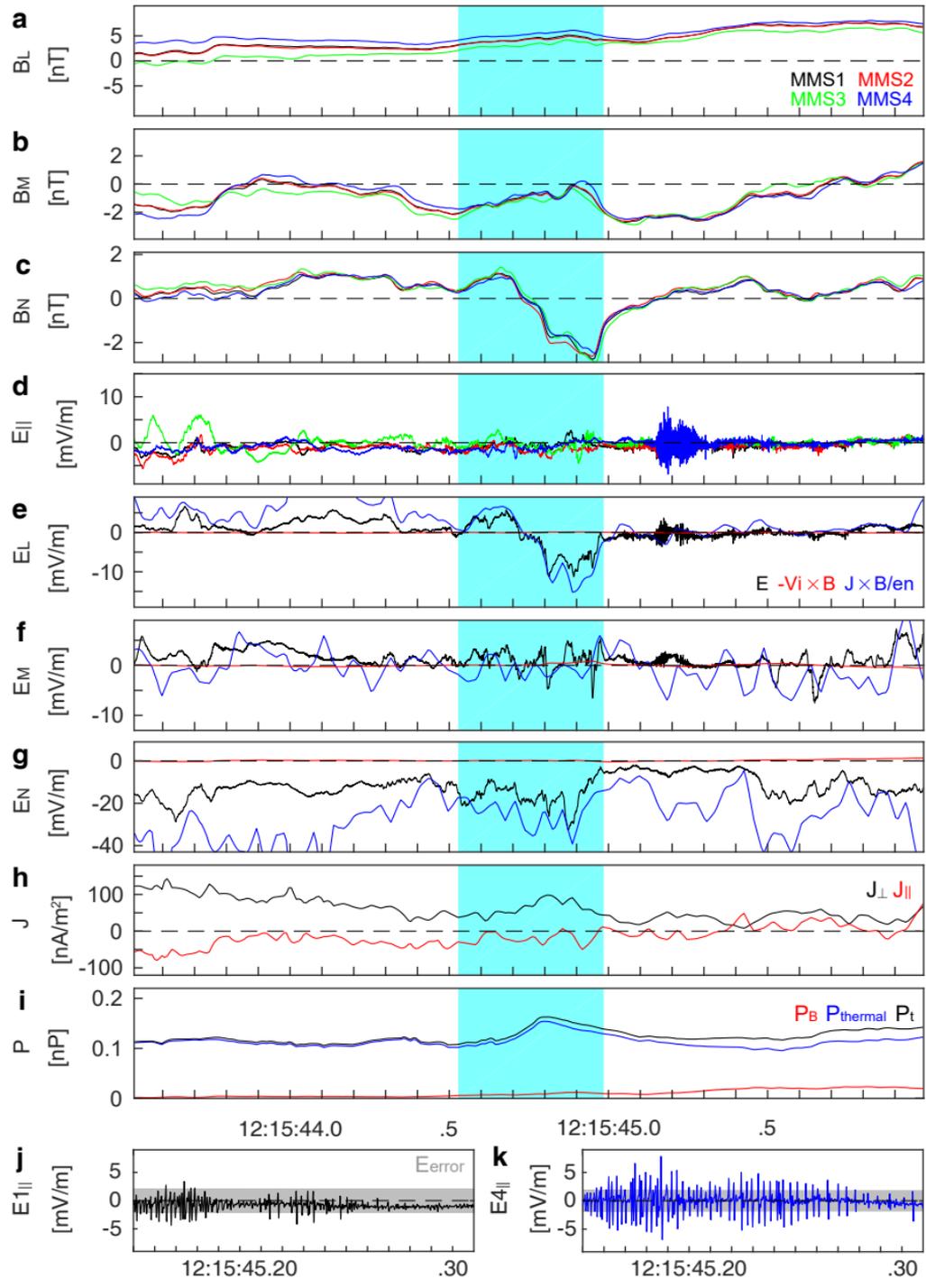